\documentclass{aa}  

\usepackage{natbib}
\bibpunct{(}{)}{;}{a}{}{,}

\usepackage[usenames,dvipsnames]{color}

\newcommand{\appropto}{\mathrel{\vcenter{
			\offinterlineskip\halign{\hfil$##$\cr
				\propto\cr\noalign{\kern2pt}\sim\cr\noalign{\kern-2pt}}}}}

\usepackage{graphicx}

\usepackage{txfonts}
\definecolor{pinegreen}{RGB}{1, 121, 111} 
\definecolor{violet}{RGB}{214,39,40}
\usepackage[breaklinks=true,colorlinks,urlcolor  = blue, linkcolor=violet,citecolor=pinegreen]{hyperref} 
\usepackage{amsmath}

 \defcitealias{Temmink2023}{T23}

\begin{document}

   \title{Good things come to those who wait}

   \subtitle{Watching donor stars evolve towards a mass-transfer instability}

   \author{     
            K. D. Temmink
            \inst{1}
            \fnmsep\thanks{e-mail: \href{mailto:Karel.Temmink@ru.nl}{Karel.Temmink@ru.nl} }
        \and
            O. R. Pols
            \inst{1}
        \and
            S. Justham
            \inst{2}
        \and
        	N. Blagorodnova
        	\inst{3,4,5}
          }

   \institute{  
            Department of Astrophysics/IMAPP, Radboud University Nijmegen, P.O. Box 9010, 6500 GL    Nijmegen, The Netherlands
        \and 
            Max-Planck-Institut f\"{u}r Astrophysik, Karl-Schwarzschild-Stra{\ss}e 1, 85748 Garching, Germany
        \and 
        	Institut de Ciències del Cosmos, Universitat de Barcelona, c. Martí i Franquès, 1, 08028, Barcelona, Spain
       	\and
       		Departament de Física Quàntica i Astrofísica, Universitat de Barcelona, c. Martí i Franquès, 1, 08028, Barcelona,
       		Spain
      	\and
      		Institut d’Estudis Espacials de Catalunya, c. Gran Capità, 2-4, 08034, Barcelona, Spain
             }

   \date{Received Month day, year; accepted Month day, year}

  \abstract
   {Unstable mass transfer in binary systems can lead to transients such as luminous red novae (LRNe). Observations of such transients are valuable for understanding and testing theoretical models of mass transfer. For donor stars in the Hertzsprung gap, there can be a long-lasting phase of mass-transfer evolution before instability sets in. Only few case studies of such delayed dynamical instability (DDI) mass transfer exist. None consider the full extent of the pre-instability evolution and the effects thereof on the observable properties of a binary.}
   {We systematically analyse detailed models of stable and unstable mass transfer for Hertzsprung-gap donors. We focus on identifying observable evolutionary features that are characteristic of ultimately unstable mass transfer and not found in stable mass-transfer binaries.}
   {Our binary evolution models, calculated with the MESA code, cover initial donor masses between $2.5 M_{\odot}$ and $10 M_{\odot}$ and initial accretor-to-donor mass ratios between $0.1$ and $1$.}
   {We find that the pre-instability evolution is qualitatively the same for all DDI donor stars, consisting of a long slow dimming phase followed by a shorter phase of rapid brightening. This latter phase is powered by recombination of hydrogen and is accompanied by a strong increase in effective temperature, which does not occur in stable mass-transfer binaries. We estimate that a significant fraction of the rapid brighteners should be detectable by Gaia throughout the Galaxy. We model the progenitors of LRNe M31-2015 and V838 Mon, find a higher initial donor mass for M31 2015 than past estimates, and propose a new scenario for V838 Mon in which the known tertiary star dominates the pre-outburst photometry and the outburst results from the DDI of a more massive primary star.
   }
  {This work provides a more comprehensive framework linking theory to observations of transients and enables improved classification and prediction of mass-transfer events.}


   \maketitle

\section{Introduction}

Astrophysical transients serve as valuable probes into the dynamic and sometimes violent processes that shape stellar evolution. Among these, luminous red novae (LRNe) are characterized by their intermediate luminosities (between those of novae and supernovae), prolonged evolution in the optical (months to years) and usually two-peaked lightcurve. An initial blue peak is followed by a prolonged redder plateau or peak powered by shocks and recombination \citep{Ivanova2013a, Metzger2017, Chen2024a, Hatfull2025} which ends with rapidly cooling ejecta \citep{Blagorodnova2017}. LRNe are believed to be a consequence of unstable mass transfer in binary star systems \citep[see e.g.][for the well-known case of V1309 Sco]{Tylenda2011}, culminating in a stellar merger or common-envelope (CE) ejection. We note that CE and merger events might also manifest as highly reddened IR-only transients \citep[see e.g.][]{Smith2016, Blagorodnova2017,Kasliwal2017,Jencson2019, Blagorodnova2021}. LRN events, their archival progenitor stars and their remnants have been observed in the Milky Way: V1309 Sco \citep{Mason2010,Tylenda2011}, V4332 Sgr \citep{Martini1999}, V838 Mon \citep{Munari2002,Bond2002}; in M31 \citep{Rich1989,Williams2015,MacLeod2017,Blagorodnova2020,Pastorello2021}, and several dozen events were also identified in other nearby galaxies, \citep[e.g. see][and references therein]{Pastorello2019a,Blagorodnova2021,Reguitti2025}.

Observations of LRNe and the archival identification of their progenitors open the possibility to directly constrain a major uncertainty in our understanding of binary-star evolution. There remains a considerable lack of certitude regarding which Roche-lobe overflow phases are unstable \citep[][hereafter T23, and references therein]{Temmink2023}. Whether a phase of Roche-lobe overflow (RLOF) in a binary system is stable or unstable is expected to lead to radically different outcomes for that phase. Additionally, even when a particular phase of mass transfer is expected to be unstable, our ability to accurately model the evolutionary outcome remains limited (\citealt{Ivanova2013}). More accurately modelling the LRN event itself and the preceding evolutionary phase would increase our understanding of the potential outcomes of unstable mass transfer.

 In a significant part of the parameter space, interacting isolated binary stars reach RLOF when the donor star is expanding across the Hertzsprung gap (HG). For such donor stars, even if the mass-transfer phase is eventually unstable, RLOF can continue for a thermal timescale before the instability \citep{Hjellming1987} and would hence typically appear stable over human timescales. During the long phase of dynamically stable RLOF before the instability, the appearance of the donor star can change significantly \citep[e.g.][]{Han2006}. This has been studied in the particular context of AT 2018bwo in NGC 45 by \cite{Blagorodnova2021}, who emphasized that using single-star models to interpret the progenitor of LRNe can be highly misleading. However, the pre-instability change in the appearance of LRN progenitors has not been more broadly systematically investigated.

In this work, we investigate the stellar physics of the donor star during the accelerating evolution towards a mass transfer instability. We seek to provide a systematic synthetic catalogue of progenitor models and their pre-transient evolution, based on calculations performed by \citetalias{Temmink2023}. This catalogue could help pinpoint potential progenitor stars and aid the interpretation of associated outbursts. Additionally, we hope that our calculations will help in the identification of binary systems which are evolving towards a mass transfer instability, ahead of the outburst.

 In Section \ref{sec:method} we describe our numerical setup and model selection. We present our main results in Section \ref{sec:pre_DDI_evolution} and discuss limitations of our work in Section \ref{sec:discussion}. In Section \ref{sec:implications_applications} we discuss implications of our results with respect to existing photometric observations and potential applications to future observations. Our conclusions are summarized in Section \ref{sec:conclusion}.

\section{Numerical methods and physical assumptions}
\label{sec:method}

We construct our synthetic catalogue from the calculations of \citetalias{Temmink2023}. Details of the numerical setup and physical assumptions can be found in \citetalias{Temmink2023}, but we briefly summarize the aspects most relevant for this work here. The binary evolution models were constructed with the \texttt{MESA} code \citep[version \texttt{r12115};][]{Paxton2011, Paxton2013, Paxton2015, Paxton2018, Paxton2019}. Assuming a metallicity of $Z = 0.02$, mass transfer sequences were calculated for non-rotating stars with masses $1-8\; M_{\odot}$. The mass transfer was taken to be fully conservative, i.e. all mass transferred was accreted by the companion star\footnote{However, some inadvertent mass and angular momentum loss does occur in these models, but should only have a marginal effect on the evolution of DDI binaries. For details, we refer the reader to \cite{Temmink2025}.}. We discuss the effects of non-conservative mass transfer in Section \ref{sec:discussion_conservative}. The accretor was treated as a point mass and its evolution was not followed. Lastly, the response of the donor stars was modelled in detail, including the effects of local thermal relaxation in the outer layers.

In this work, we focus on post main-sequence stars with initial masses in the range of $2.5-10\;M_{\odot}$ that have radiative envelopes at the onset of mass transfer. The additional models calculated for this work have a numerical setup that is identical to the one in \citetalias{Temmink2023}, ensuring internal consistency of our models. We include in our DDI model set those binary systems that would experience a delayed dynamical instability based on the critical mass ratio for quasi-adiabatic evolution \citepalias[see][]{Temmink2023}. That is, we select the binary systems with $q \equiv M_{\rm a}/M_{\rm d} < q_{\rm qad}$. Typically, this results in the following selection of mass ratios at the onset of mass transfer: $q \in \{ 0.1, 0.15, 0.2, 0.25 \}$. The binaries with initially larger values of $q$ (up to and including $q = 1$) form the stable mass-transfer model set. %

\section{Evolution towards the mass transfer instability}
\label{sec:pre_DDI_evolution}

In this section, we present our main results. In Section \ref{sec:representative_evolution}, we describe typical aspects of the pre-DDI evolution in the HRD, and how these differ from a system undergoing stable mass transfer. A more detailed consideration of pre-DDI evolution, which focuses on the evolution of the internal stellar structure, is given in Section \ref{sec:evolution_details}. We discuss how those results vary through the parameter space we studied in Section \ref{sec:parameter_space}.

\subsection{Representative observable evolution of a DDI binary}
\label{sec:representative_evolution}

\begin{figure*}[h!]
	\centering
	\includegraphics[width=\hsize]{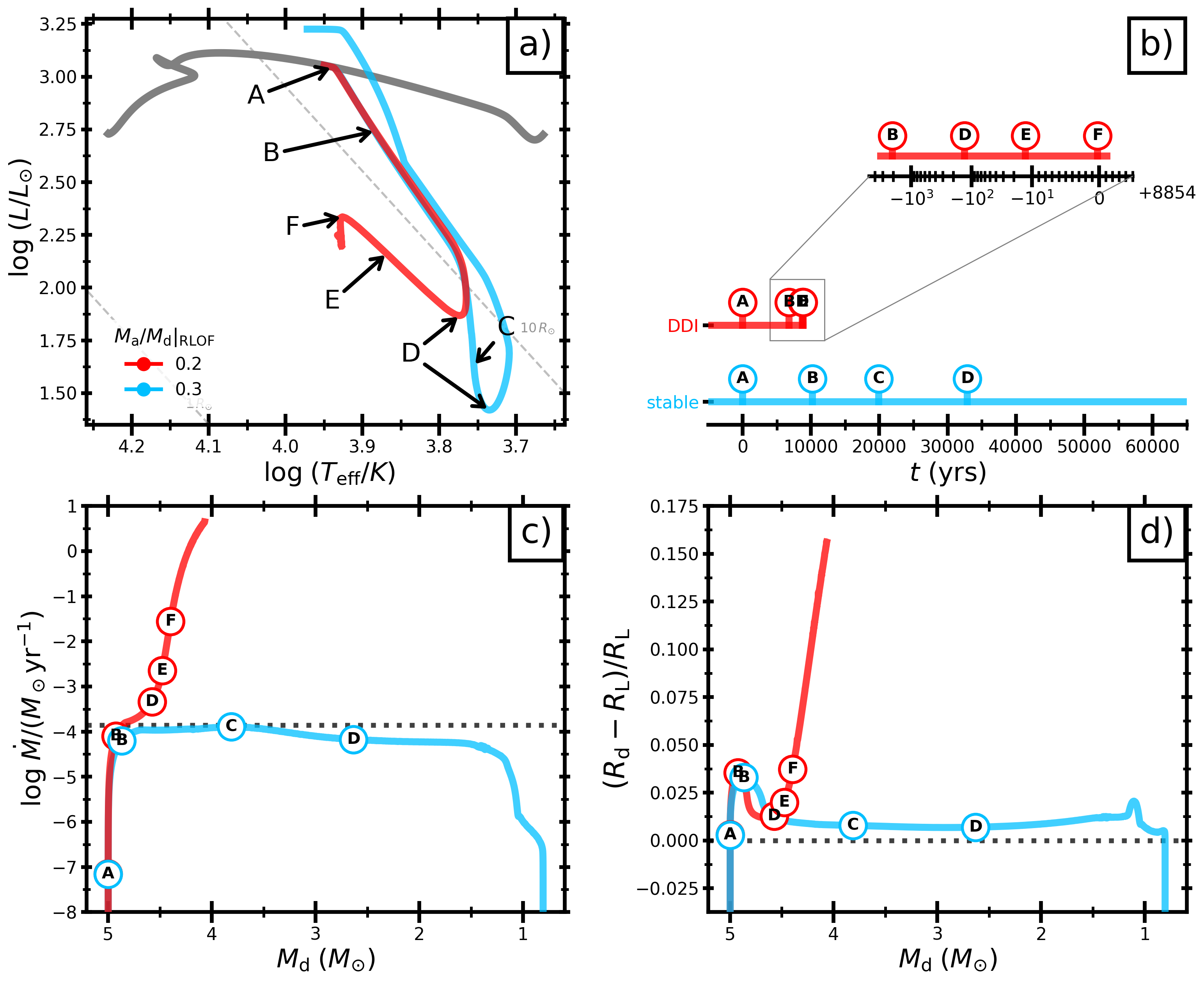}
	\caption{Representative evolution of two $5\;M_{\odot}$ donor stars, one that will experience a DDI, and one that will not. The donor stars fill their Roche lobes at the same moment in their detached evolution (at a size of $15\;R_{\odot}$), but lose mass in different binary configurations. The DDI system, shown in red, contains an initially $1\;M_{\odot}$ companion star, whilst the stable mass transfer system, shown in blue, starts with an initially $1.5\;M_{\odot}$ companion. Points of interest are labelled with letters, which are explained in the text and Table \ref{tab:points_of_interest}. In the lower left panel, the dotted line shows the mass transfer rate based on the Kelvin-Helmholtz thermal timescale at point A.}
	\label{fig:HRD_labels}
\end{figure*}

\begin{table}[h!]
	\caption{Codification for key moments in the evolution of mass-losing stars}    
	\label{tab:points_of_interest}     
	\centering                          
	\begin{tabular}{l|l}        
		\hline\hline                 
		label & description \\   
		\hline                        
		A     & Onset of RLOF.                                                   \\
		B     & \vtop{\hbox{\strut Moment of maximum Roche-lobe over-extension.}\hbox{\strut (before point E)}}           \\
		C     & \vtop{\hbox{\strut Moment where $\dot{M}_{\rm d}$ is maximum.}\hbox{\strut (non-DDI models only)}}         \\
		D     & Luminosity minimum. \\
		E     & Moment the qad criterion \citepalias[see][]{Temmink2023} is reached.    \\ 
		F     & \vtop{\hbox{\strut Local luminosity maximum during $\dot{M}_{\rm d}$ increase.}\hbox{\strut Near-adiabatic stratification of the envelope.}}\\
		\hline                                   
	\end{tabular}
\end{table}

Here, we describe the general aspects of evolution in the HRD leading up to a DDI. Many aspects of mass transfer in binaries have been studied and documented extensively, especially regarding binaries undergoing stable mass transfer. We briefly summarize those points, but focus here on those aspects pertinent to pre-DDI evolution. For this purpose, and in order to highlight the difference in evolution between stable mass transfer and a DDI, we present the evolution of two representative binary systems in Figure \ref{fig:HRD_labels}. The $5\;M_{\odot}$ donor stars in these binaries are initially identical and they fill their Roche lobes at the same moment in their evolution (at a radius of about $15\;R_{\odot}$). The only difference is that the DDI system (shown in red) begins its evolution with a companion star of $M_{\rm a}= 1 M_{\odot}$ (such that $M_{\rm a}/M_{\rm d} = 0.2 < q_{\rm qad}$) and at a corresponding initial period of $7.1$ days, whilst the stable mass-transfer system (shown in blue) starts with a companion star of $M_{\rm a}= 1.5 M_{\odot}$  (such that $M_{\rm a}/M_{\rm d} = 0.3 > q_{\rm qad}$) with a corresponding initial period of $7.6$ days. Throughout this work, we will refer to several key moments of binary evolution that occur in all the binaries studied here. These moments are codified in Table \ref{tab:points_of_interest}.

\begin{figure*}[h!]
	\centering
	\includegraphics[width=\hsize]{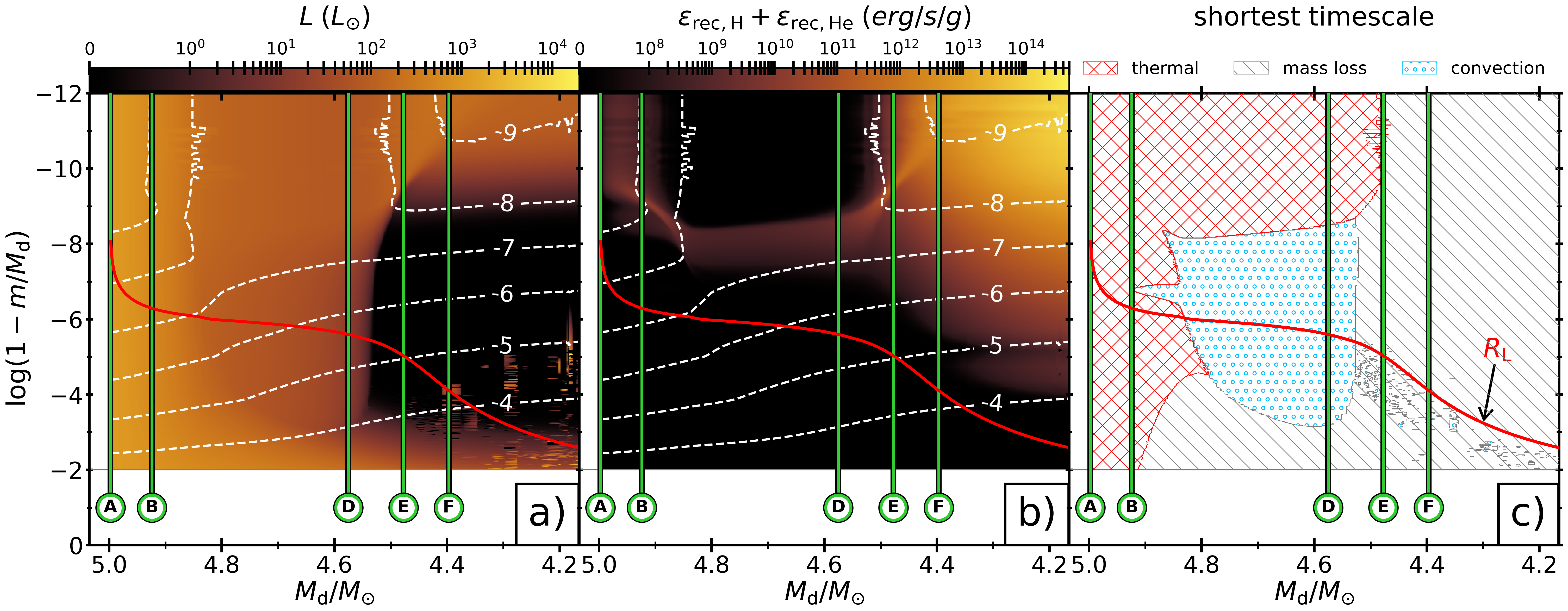}
	\caption{Structural changes in the donor star of a DDI binary system, shown for the same representative DDI system as in Figure \ref{fig:HRD_labels}: a donor star with a ZAMS mass of $5\;M_{\odot}$ that fills its Roche lobe at a size of $14.3\;R_{\odot}$ and loses mass to an initially $1\;M_{\odot}$ companion. From left to right, the panels show how the internal luminosity, specific energy generation due to H and He recombination and shortest timescale vary throughout the donor star during the mass transfer. For visual clarity, negative values (which have a minimum of about $-2\cdot 10^{6}$ ergs/s/g) are not shown in panel b. The total mass of the donor star $M_{\rm d}$ decreases from left to right along the horizontal axis. The vertical axis shows the amount of mass exterior to a shell at mass coordinate $m$, relative to the total donor mass. In this coordinate system, the core of the star lies towards the bottom of the figure, and the envelope towards the top. Lines of constant $\log\;\rho$ are drawn in white and dashed, with the corresponding values in cgs units indicated on the right hand side of each panel. The solid red line shows the location of the Roche lobe equivalent volume radius inside the star. The vertical green lines and associated letters correspond to the moments shown in Figure \ref{fig:HRD_labels} and described in Table \ref{tab:points_of_interest} and the text.}
	\label{fig:DDI_structure_2D}
\end{figure*}

Our calculations start before the donor stars have filled their Roche lobes, and they thus initially evolve nearly parallel to the corresponding single star track (shown in Fig.\,\ref{fig:HRD_labels} as a grey line). At the point labelled A in Table \ref{tab:points_of_interest} and Figure \ref{fig:HRD_labels}, the donor star fills its Roche lobe and the mass transfer rate increases sharply. In reaction to the loss of mass, both the Roche-lobe radius and the stellar radius decrease. Formerly interior layers are lifted up to the surface, which uses energy \citep[see e.g.][]{Webbink1976, Webbink1977}, reducing the star's luminosity. As these layers expand and cool, the star moves downwards and to the right through the HRD.

As the mass-transfer rate increases during the phase following point A, the donor star extends increasingly far beyond its Roche lobe. As Figure \ref{fig:HRD_labels}d) shows, this occurs in both the DDI binary and the stable mass transfer binary, and is not a sign of the mass transfer turning unstable. Due to a restructuring of the envelope (see Section \ref{sec:evolution_details}), the mass-transfer rate can eventually be maintained by a smaller excess radius and a turning point is reached in the extent of overflow at point B, about eight thousand years after the start of RLOF for the DDI binary and approximately ten thousand years for the stable mass-transfer binary.

After point B, one can start to see the first marked differences between the evolution of a DDI binary and that of a stable mass transfer system. In the stable system, the mass-transfer rate increases slightly over a period of tens of thousands of years, until the maximum $\dot{M}_{\rm d}$ (which is set by the evelope thermal timescale) is reached at point C. In the DDI binary, however, the mass-transfer rate increases more sharply, as shown in Figure \ref{fig:HRD_labels}c), to a rate much higher than that corresponding to the global thermal timescale, even through the extent of RLOF is decreasing.

Eventually, both binaries reach a turning point in the HRD at point D, after which the luminosity in both donor stars begins to increase. However, the condition in which the donor stars reach point D, as well as the reason for the turnaround and whether the donor turns to higher or lower effective temperatures, are different between the stable and DDI system.
The stable system arrives at point D with a mass of 2.65$M_{\odot}$ about 33000 years after the onset of RLOF. The DDI system, however, reaches point D at a mass of 4.575$M_{\odot}$ about 8700 years after the onset of RLOF.

In the stable system, the mass-transfer rate continues to decrease over tens of thousands of years. The luminosity increases as the donor gradually starts to recover its gravothermal equilibrium and moves to lower effective temperatures. After the upturn towards the right in the HRD, the donor star eventually moves back to the upper left after most of its envelope is stripped.

The DDI donor also evolves to higher surface luminosity. However, in this case the increase is due to the energy released by the recombination of ionized hydrogen in the ever more rapidly uplifted outer layers (see Section \ref{sec:evolution_details}). At the same time, the size of donor star still roughly follows that of the shrinking Roche lobe (see Figure \ref{fig:HRD_labels}d). As a result, the donor star evolves to higher effective temperatures.

As the mass transfer rate ever increases in the DDI system, eventually at point E the mass-transfer rate exceeds the critical thermal rate \citepalias[for a detailed explanation, see][]{Temmink2023} and no part in the envelope of the donor stars can thermally readjust on a timescale shorter than the mass-loss timescale. 

When the DDI donor star approaches point F, the envelope is nearly adiabatically stratified. Subsequently, the luminosity of the donor star  drops rapidly at nearly constant effective temperature, and the star traces another hook through the HRD at point F. During the evolution between points D and F, the donor star becomes brighter by a factor of about three in roughly a century. Our calculation stops soon after F owing to numerical issues, but in any case we expect our 1D hydrostatic model to become increasingly unreliable beyond F.

\subsection{Detailed structural pre-DDI evolution}
\label{sec:evolution_details}

\begin{figure}[h!]
	\centering
	\includegraphics[width=\hsize]{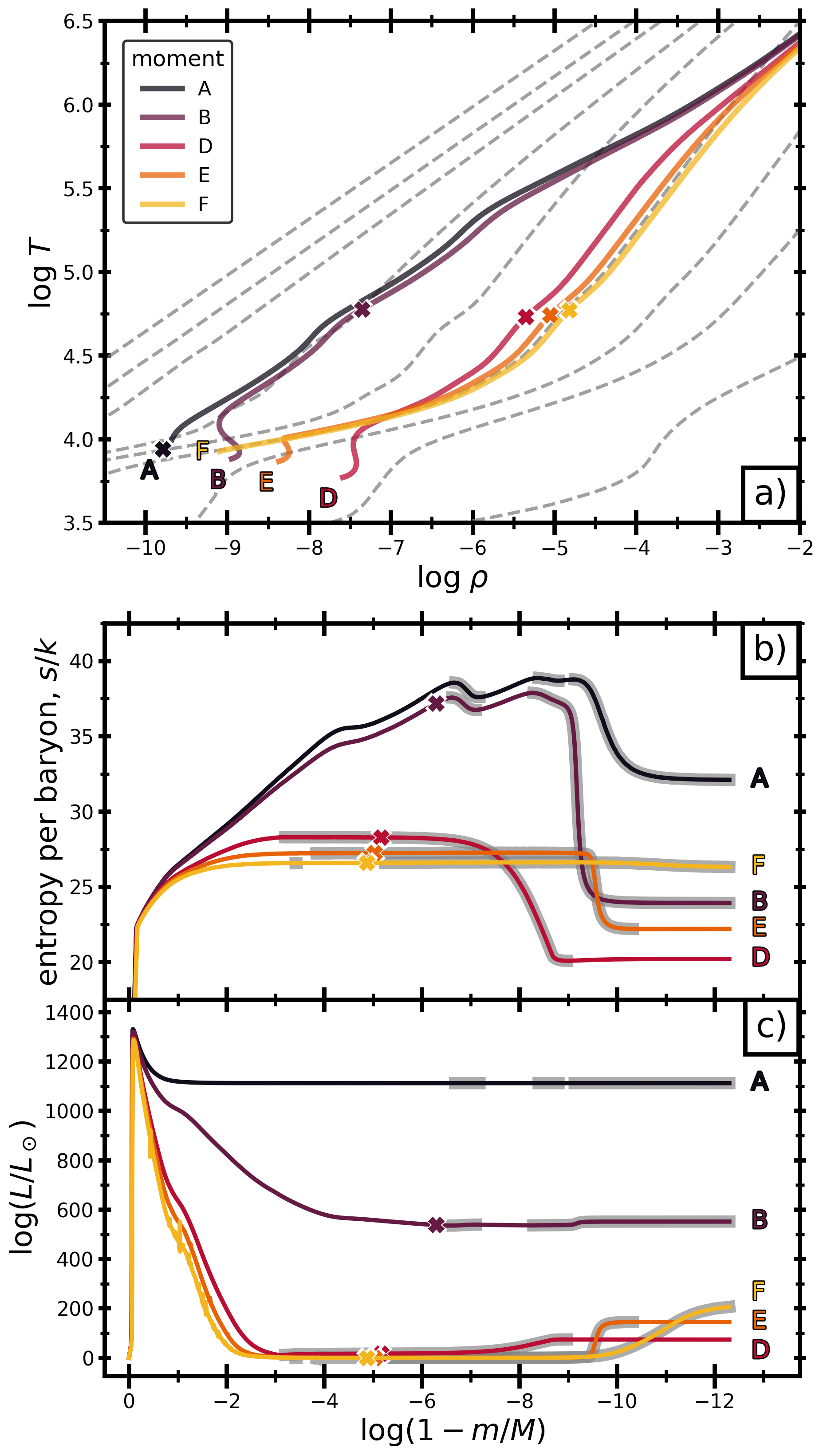}
	\caption{Structural changes in the donor star of the same representative DDI system as in Figures \ref{fig:HRD_labels} and \ref{fig:DDI_structure_2D}. Panel a shows the structure of the donor star in the $\log \rho - \log T$ (in cgs units) plane at different moments in the evolution. Panels b and c show how, respectively, the specific entropy (per baryon) and internal luminosity vary throughout the donor star, using the fractional exterior mass as horizontal coordinate. In panel a, gray dashed lines correspond to adiabats. In panels b and c, gray shading indicates where the star is convective.   In all panels, the lines are coloured and labelled according to the moments in the legend (see also Table \ref{tab:points_of_interest}). Crosses indicate the location of the Roche lobe equivalent volume radius inside the star. }
	\label{fig:DDI_structure_lines}
\end{figure}

In this section, we address in more detail the physical processes that produce the observable behaviour presented in Section \ref{sec:representative_evolution}, focusing on the phases beyond point D which are pertinent to DDI evolution and differ most from the evolution of stable mass-transfer systems.

During the mass transfer evolution, a DDI donor star will generally shrink in overall physical size. However, individual layers within the star expand in reaction to the sudden loss of mass and their structure changes. Many aspects of the observable pre-DDI evolution are determined by these structural changes in the outermost layers. Precisely how and on what timescale these layers expand, in turn, affects the mass transfer in a positive feedback loop. Figures \ref{fig:DDI_structure_2D} and \ref{fig:DDI_structure_lines} shows how several important aspects of the interior structure evolve during the mass transfer. Of particular interest is the subsurface region where the temperature gradient is steeper than in the surrounding areas where it is nearly adiabatic. This results in the negative entropy gradients seen in Figure \ref{fig:DDI_structure_lines}b. In that region, known as the superadiabatic (SAd) zone (see e.g. \citetalias{Temmink2023} and references therein), the comparatively short local thermal timescales allow that part of the star to restructure itself during a large part of the mass-transfer evolution.

The evolution of the structure in the SAd region is linked to the evolution of the  donor star across the HRD. Prior to the luminosity minimum at point D, the SAd region moves inwards in both absolute and relative mass coordinate as the~envelope~restructures in reaction to the loss of mass. As the donor star evolves through point D, however, the SAd region shrinks in mass and moves outwards towards the surface. This can be seen most clearly by examining the entropy profiles at point D and subsequent points in Figure \ref{fig:DDI_structure_lines}b. During this phase, the outermost subsurface layers have less and less time to thermally restructure and the temperature there increases as the structure is forced to follow an adiabat (see Figure \ref{fig:DDI_structure_lines}a). Consequently, the donor stars reach a turning point in the HRD at point D, after which the effective temperature increases. 

The ever more rapid expansion of the subsurface layers results in a substantial decrease in internal luminosity (the total rate of energy transported) in most of the subsurface layers, as Figures \ref{fig:DDI_structure_2D}a and \ref{fig:DDI_structure_lines}c show. More explicitly, the luminosity is nearly zero throughout most of these layers, as nearly all the luminosity from the inner regions is used to facilitate the expansion. In the outermost parts of the subsurface layers, however, the expanding layers have cooled sufficiently for the previously ionised hydrogen to recombine. The rate of recombination, and consequently the rate of change in internal energy, is roughly proportionate to the rate of mass loss and thus ever increases throughout the DDI evolution. Figure \ref{fig:DDI_structure_2D}b shows a clear substantial increase in the recombination energy generation rate in the outermost layers starting at point D. The luminosity arising from this rapid recombination eventually becomes comparable to the luminosity absorbed in the expansion of the star by point D. As a result, the surface luminosity of the star, which had been decreasing, now strongly increases and the star moves upwards and to the left in the HRD.

At point E, the mass-transfer rate has grown to such an extent that no part in the envelope of the donor stars can thermally readjust on a timescale shorter than the mass-loss timescale. This is demonstrated in Fig \ref{fig:DDI_structure_2D}c, which shows the shortest timescale (set by either local thermal relaxation, convective turnover, or mass loss) throughout the star. This is the moment where the mass-transfer rate exceeds the critical thermal rate as defined in \citetalias[][]{Temmink2023}. It was found in \citetalias{Temmink2023} that this moment marks a transition to quasi-adiabatic evolution, which occurs in all systems that will experience a mass-transfer instability. 

When the donor star approaches point F, the envelope is nearly adiabatically stratified (see Figs. \ref{fig:DDI_structure_lines}a and b). The superadiabaticity $\nabla - \nabla_{\rm ad}$, with $\nabla \equiv \partial \log T/\partial \log P$, in the outermost layers is typically only about 0.05 (for comparison, in a typical giant star the superadiabaticity can reach values in excess of 1.75). As the response of the donor star to the loss of mass becomes very close to adiabatic, nearly all the energy coming from the release of internal energy by cooling and recombination goes into the work done by expanding the layers inside the envelope. As a result, the luminosity of the donor star drops rapidly at nearly constant effective temperature, and the star traces another hook through the HRD at point F. Thereafter, the effective temperature remains roughly constant as the structure of the outer layers of the donor star keeps following the same adiabat. 

\subsection{Variation of observable properties with initial parameters}
\label{sec:parameter_space}

\begin{figure*}[h!]
	\centering
	\includegraphics[width=\hsize]{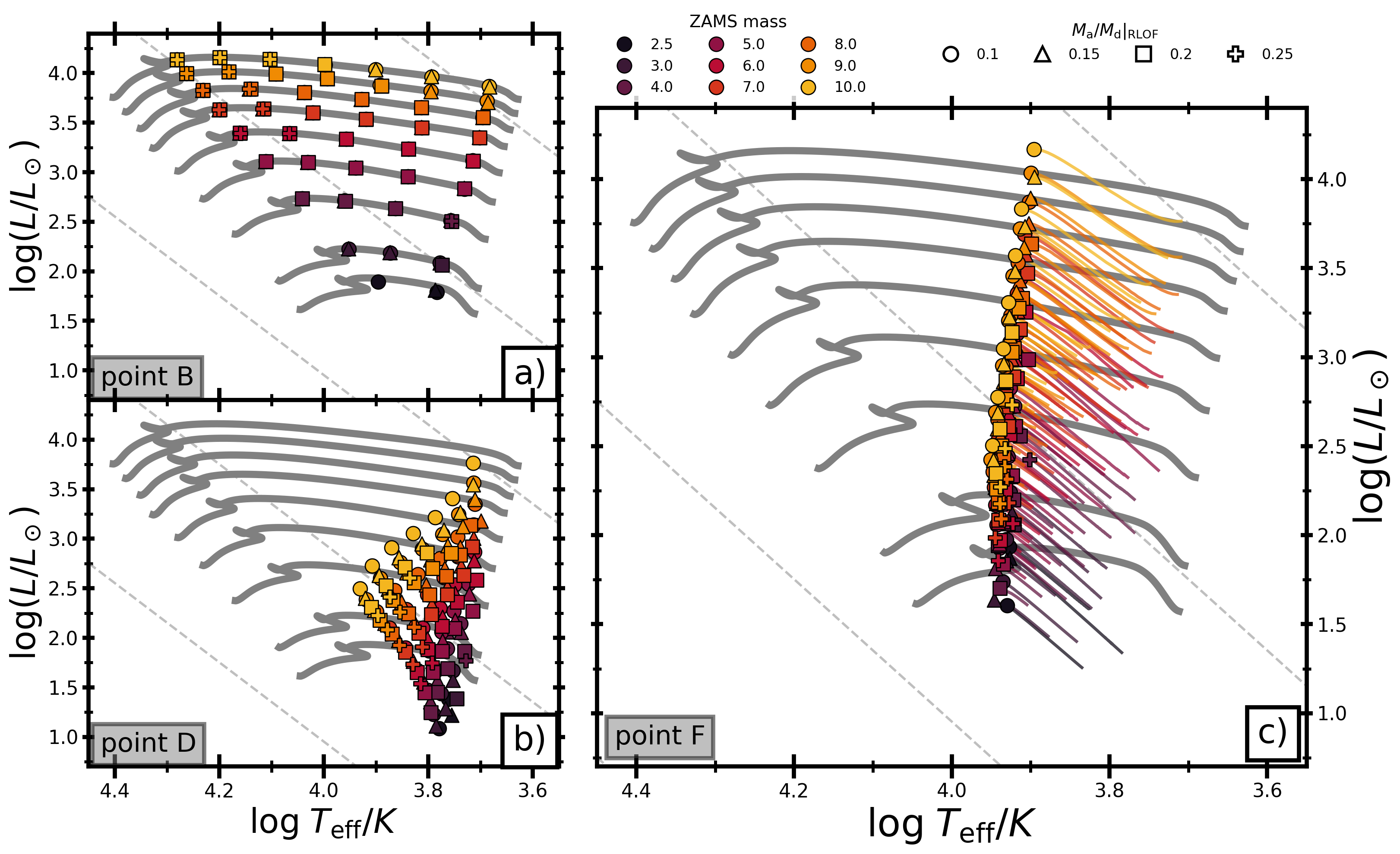}
	\caption{HRD locations of all our models at point B (the onset of RLOF; panel a), point D (the luminosity minimum; panel b) and point F (near-complete adiabatic stratification of the envelope; panel c). In all panels, the data points are coloured according to the corresponding initial donor mass and the marker type reflects the initial mass ratio, as indicated in the legend. Panel c furthermore shows the HRD tracks of the donor stars during the 10 years preceding point F. }
	\label{fig:HRD_BDF}
\end{figure*}

The pre-instability evolution of all the DDI binaries in our calculations is qualitatively the same as in the representative example discussed in Sections \ref{sec:representative_evolution} and \ref{sec:evolution_details}. However, we find a few systems that have $q < q_{\rm qad}$ and temporarily evolve effectively adiabatically but nonetheless do not run away to dynamical-timescale mass transfer. These systems are all early-HG stars when they fill their Roche lobes with $q=0.25,0.3$. We do not consider these systems proper DDI binaries and will not discuss them further in this work, but we cannot exclude that these systems would produce an outburst due to the response of the companion. The behaviour of these particular few systems has very little influence on the critical mass ratios derived in \citetalias{Temmink2023}.

For the DDI binaries, there are important quantitative variations in the observable evolution that depend on the different starting configurations of the binaries. Figure \ref{fig:HRD_BDF} offers a summary of the HRD evolution of all our models, showing their luminosities and effective temperatures at points B, D and F. 

The HRD tracks of the donor stars naturally start at lower effective temperatures for initially more evolved donors and at larger luminosities for initially more massive donors, as shown in Figure \ref{fig:HRD_BDF}a. Consequently, the effective temperature at point D is also lower for initially more evolved stars, whilst the luminosity at that point increases strongly with mass, as shown in Figure \ref{fig:HRD_BDF}b. For a given initial donor mass and radius, the HRD location of point D depends strongly on the initial mass ratio of the binary. The mass-transfer rate increases faster for more extreme initial mass ratios, hence point D is reached earlier in the evolution of such systems. As a result, the donor stars with lower initial $M_{\rm a}/M_{\rm d}$ have shrunk less by that time and arrive at point D with a higher luminosity ($L_{\rm D}$; see Figure \ref{fig:HRD_BDF}b). For a given initial mass ratio, $L_{\rm D}$ typically increases with the initial period, roughly following the trend $\log L_{\rm D} \appropto 0.8\log P_{\rm RLOF}$. Likewise, $T_{\rm eff, D}$ is larger for smaller $M_{\rm a}/M_{\rm d}$, though the dependence of $T_{\rm eff, D}$ on the mass ratio is weak, except for the binaries with initial periods $P_{\rm i} < 10$ days.

After reaching point D, all the donor stars start to increase in brightness, as explained in Sections \ref{sec:representative_evolution} and \ref{sec:evolution_details}. Between D and F, the luminosity of the donor stars increases by up to a factor of 5. The increase in luminosity is larger for initially wider and more massive binaries, but is nearly independent of the initial mass ratio. The time it takes the donor stars to evolve between D and F ($t_{\rm DF}$) depends strongly on the initial mass ratio of the binaries, but typically ranges between 10 and 500 years, with the shorter timescales applying to systems with initially more extreme mass ratios.

Despite their very different starting points, our models reach point F with remarkably similar effective temperatures around $\log T_{\rm eff}/K = 3.9$ (see Figure \ref{fig:HRD_BDF}c). The full range is between $\log T_{\rm eff}/K = 3.85$ and $\log T_{\rm eff}/K = 3.95$. As explained in Section \ref{sec:evolution_details}, the stratification of the outer layers is nearly adiabatic at point F. In regions where hydrogen is partially ionized, the adiabats in $\log \rho-\log T$ plane flatten out near the surface, and adiabats that were spaced farther apart at higher $\log \rho$ and $\log T$ converge near $\log T/K = 3.9$. The range in surface luminosity spanned by our models at point F is substantial, even for models with the same initial donor mass (Figure \ref{fig:HRD_BDF}c). The spread is relatively larger for initially more massive donor stars, which generally have larger surface luminosities at this point.

By the time that the evolution starts to deviate from that of stable mass-transfer binaries (from point D onwards), many properties of the binaries have changed significantly from their values at the start of RLOF. The loss of mass, smaller radius and restructuring of the envelope has consequences for the binding energy of the envelope and common envelope evolution. This is discussed in more detail in Appendix \ref{sec:CEE}.

\section{Limitations of our models}

\label{sec:discussion}

This work is based on calculations performed for \citetalias{Temmink2023}. As such, the uncertainties, limitations and caveats --both numerical and physical-- presented there apply here, too. For a detailed description of those, we refer the reader to the discussion in \citetalias{Temmink2023}. In the following, we discuss those matters most pertinent to the context of this work.

\subsection{Evolution of the accreting companion star}
\label{sec:discussion_accretor}

As discussed by \citetalias{Temmink2023}, the evolution of the accretor was ignored by treating it as an all-accreting point mass. This approach neglects the accretor's finite size, as well as structural changes which can become important when high mass transfer rates occur. By assuming that the accretor behaves as a ZAMS star in equilibrium, we can compute a conservative estimate of the accretor radius during the mass transfer. We interpolate linearly in [Fe/H] between two sets of MIST models \citep{Dotter2016, Choi2016}\footnote{Available here: \href{https://waps.cfa.harvard.edu/MIST/ }{https://waps.cfa.harvard.edu/MIST/}.} for $[\mathrm{Fe/H}] = 0$ and $[\mathrm{Fe/H}] = 0.25$, to match our $[\mathrm{Fe/H}] = 0.15$. These models have a nearly identical physical setup as our models, and are thus well-suited to obtain an appropriate ZAMS mass-radius relationship. Using this approximation, we find that none of the accretor stars fill their own Roche lobe during the entire pre-DDI evolution studied here.

However, accreting stars may experience substantial expansion if they are unable to thermally adjust quickly enough \citep[e.g.][]{Benson1970, Kippenhahn1977, Neo1977, Lau2024}. This expansion can cause the accretor star to fill its own Roche lobe, leading to a contact binary configuration. Although this scenario does not necessarily result in a common envelope, it may cause mass and angular momentum to be lost from the system through the L2 point, leading to non-conservative mass transfer (see also Section \ref{sec:discussion_conservative}). It has long been known that the critical mass ratios for which accreting stars fill their Roche lobes during MS and HG stages can be less extreme than those derived from the donor response alone \citep[see e.g.][]{Pols1994,Wellstein2001,Zhao2024}. This suggests that a contact binary and L2 mass loss could potentially occur before any instabilities due to the evolution of the donor star would occur in the binaries studied here. As a result, an observable transient may already start before before the donor star can evolve to its point F.

\subsection{The effects of non-conservative mass transfer and angular-momentum loss}
\label{sec:discussion_conservative}

\begin{figure}[t!]
	\centering
	\includegraphics[width=\hsize]{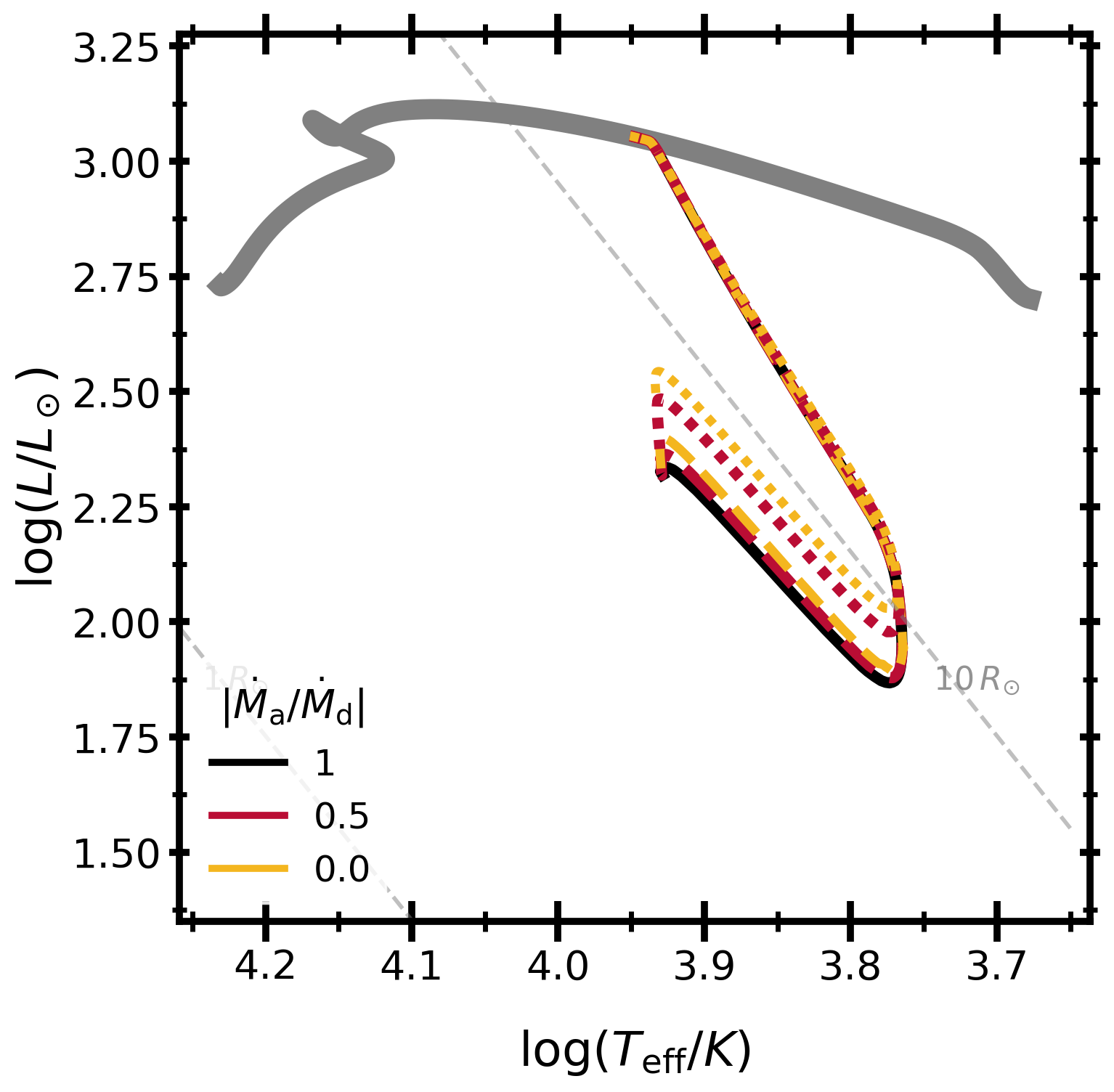}
	\caption{HRD evolution for our representative binary system (Figure \ref{fig:HRD_labels}) under different assumptions for the mass-transfer efficiency. All non-accreted matter is taken to leave the system with the specific AM of either the accretor (dashed lines) or a circumbinary ring with a radius of 1.2 times the semi-major axis (dotted lines). The line colours correspond to different mass-transfer efficiencies, as indicated in the legend.}
	\label{fig:beta_HRD}
\end{figure}

Our assumption of conservative mass transfer is a limiting case (see also \citetalias{Temmink2023}). 
Loss of mass and the associated angular momentum (AM) affects the orbital dynamics of a binary system in a variety of ways. Depending on the mode, or combination of modes, of mass loss, the effect can lead to either an increased shrinking or widening of the binary orbit compared to the fully conservative case \citep[see e.g.][]{Soberman1997}.

In order to gain understanding of the possible effects of enhanced AM loss on our main results, we have run additional simulations of our representative example DDI binary (see Section \ref{sec:representative_evolution}) and several others with enhanced mass and AM loss. Specifically, we consider cases where either half or none of the matter lost by the donor is accreted by the companion. The matter that is not accreted leaves the binary system, carrying away the specific AM of either the orbit of the accretor star or the specific AM of a circumbinary ring of radius 1.2 times the semi-major axis $a$ \citep[as appropriate for L2 outflows, see e.g.][]{Pribulla1998}. The results of this exercise are shown in Figure \ref{fig:beta_HRD}. The evolution of the donor stars in unstable mass-transfer configurations still follows the qualitative picture outlined in Section \ref{sec:representative_evolution} and all points A-F are present in the evolution. However, in binaries with enhanced mass and AM loss, point D occurs earlier and at a higher surface luminosity than it would in a fully conservative analogue. As a result, the rapid brightening stage is shifted up in the HRD as well. We find that the pre-DDI evolution is generally faster for binaries with less efficient mass transfer and hence such binaries may be more easily detected observationally. Furthermore, the mass lost from a binary system can significantly affect the photometry of a pre-outburst binary, as demonstrated for e.g. V1309 Sco \citep{Pejcha2017}.

\section{Implications for LRN outburst progenitors and Gaia observations}
\label{sec:implications_applications}

In this Section, we apply our models of pre-DDI binary evolution to the progenitors of the outburst in V838 Mon and the M31 2015 LRN outburst in order to estimate possible progenitors of said outbursts. Furthermore, we assess the implications of our models for the characterization of DDI systems using Gaia time-domain data.

\subsection{The outburst of V838 Mon}
\begin{figure}[h!]
	\centering
	\includegraphics[width=\hsize]{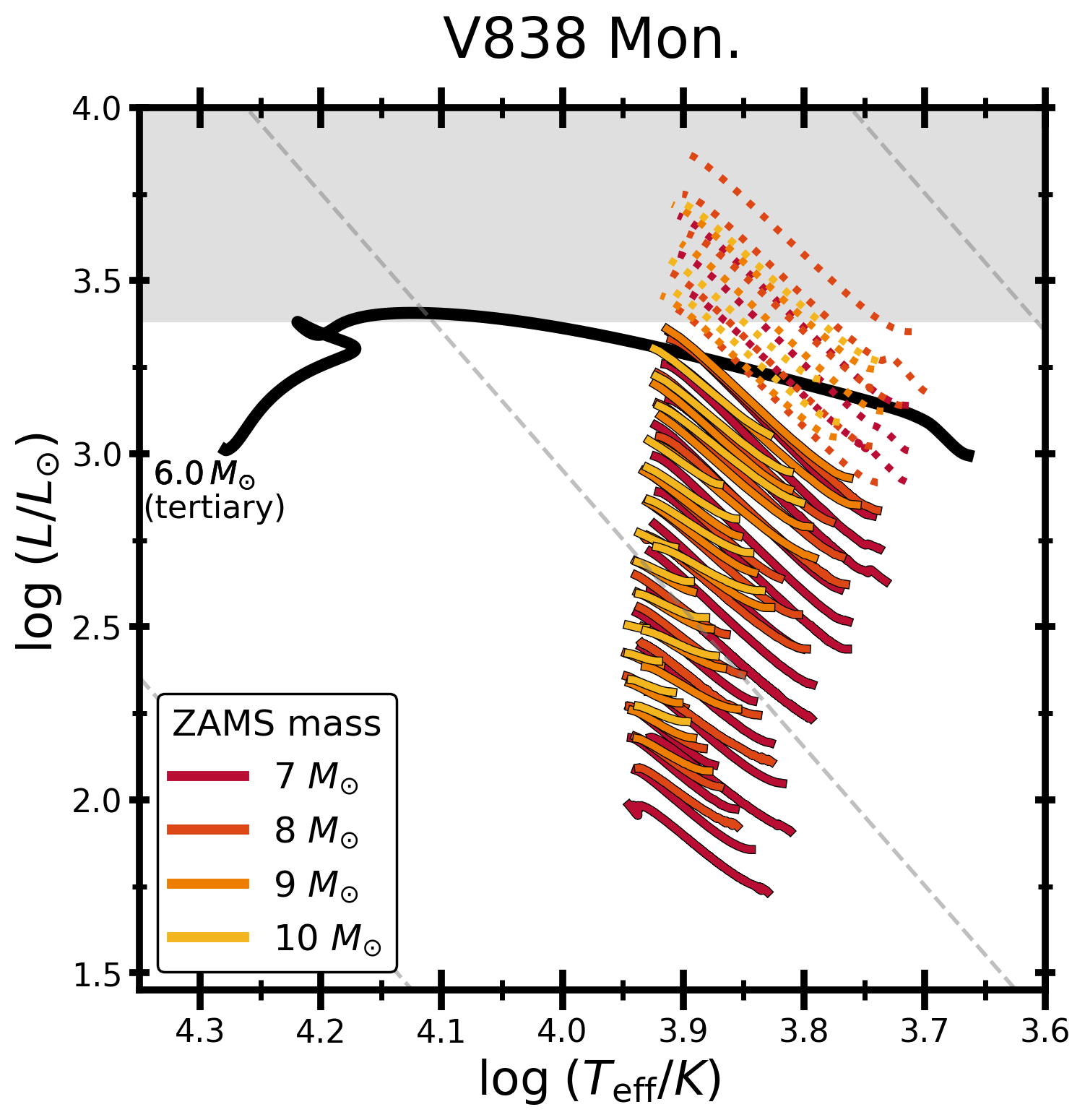}
	\caption{DDI progenitor models for the primary star in the V838 Mon merger eruption, for an assumed tertiary mass of $6\;M_{\odot}$. The thick black line shows the HRD track for a $6\;M_{\odot}$ single star. Coloured lines show the brightening phase (i.e. the evolution between points D and F) for binaries where the donor luminosity is below that of the TAMS luminosity of the tertiary. Solid lines correspond to models where this holds for the entire brightening phase, whilst dotted lines show models where this holds for only part of the brightening phase.}
	\label{fig:V838Mon}
\end{figure}

V838 Monocerotis (V838 Mon) drew significant attention from the astrophysical community following its outburst in January 2002 \citep{Brown2002}. Unlike classical novae, V838 Mon's outburst did not expel significant amounts of material at high velocities. Instead, the spectrum of the outburst evolved to resemble a cool supergiant \citep[see e.g.][]{Tylenda2005}.  It is broadly accepted that the eruption was the result of a stellar merger \citep[e.g.][]{Tylenda2006, Kaminski2021}. The stellar system which produced the outburst has a B-type companion \citep{Munari2002a}, now presumed to be a star on a wide orbit around a merged inner binary. \citet{Tylenda2005} used photometry of the pre-outburst system to constrain the primary star in the inner binary, assuming that the B-type companion is coeval with the inner binary. In particular, \citet{Tylenda2005} exclude the inner binary from containing a post-main-sequence star more luminous than the presumed tertiary, and so argue that the merger probably involved a main-sequence or pre-main-sequence star of $5-10\;M_{\odot}$ (see also \citealt{Tylenda2006}).

Here we briefly explore the possibility that the primary star in the merging inner binary was the most massive star in the system, but was nonetheless sufficiently under-luminous to remain consistent with the pre-outburst photometry. In this scenario, the tertiary companion star evolves without interacting with the merging binary system. Since the primary star in the merging inner binary is the most massive star in the triple system, it is naturally the first star to evolve to fill its Roche lobe. However, the ensuing mass-transfer is not immediately unstable and the donor star evolves for an extended period of time, as described in Section \ref{sec:pre_DDI_evolution}. In the evolution up to the luminosity minimum (point D), the primary star becomes dimmer to such an extent that its surface luminosity is well below that of its main-sequence tertiary companion. Even during the phase of rapid brightening that follows (between points D and F), the primary star's luminosity can remain below that of the tertiary.

To explore the feasibility of this scenario, we do the following for each of the relevant masses in our parameter space ($2.5 - 10 \;M_{\odot}$). For an assumed mass of the tertiary companion, $M_{\rm tert}$, we select from our models those binary systems with donor masses $M_{\rm d} > M_{\rm tert}$ whose luminosity during the rapid brightening phase is below that of the TAMS associated with the chosen tertiary mass. We find a significant parameter space of primary star progenitors for any choice of tertiary mass. We show a representative example for $M_{\rm tert} = 6\; M_{\odot}$ in Figure \ref{fig:V838Mon}. Unfortunately, due to the dearth of observational constraints on both the progenitor of the V838 Mon outburst as well as the tertiary, which is currently obscured by dust \citep{Kaminski2021}, we cannot narrow down the range of progenitor options. However, this interpretation is a qualitative change from a MS merger and is more consistent with other LRNe with known progenitors, which are typically thought to have post-MS progenitors. This exercise suggests that the scenario considered here could be a viable formation channel.

\subsection{The M31 2015 luminous red nova outburst}

\begin{figure}[h!]
	\centering
	\includegraphics[width=\hsize]{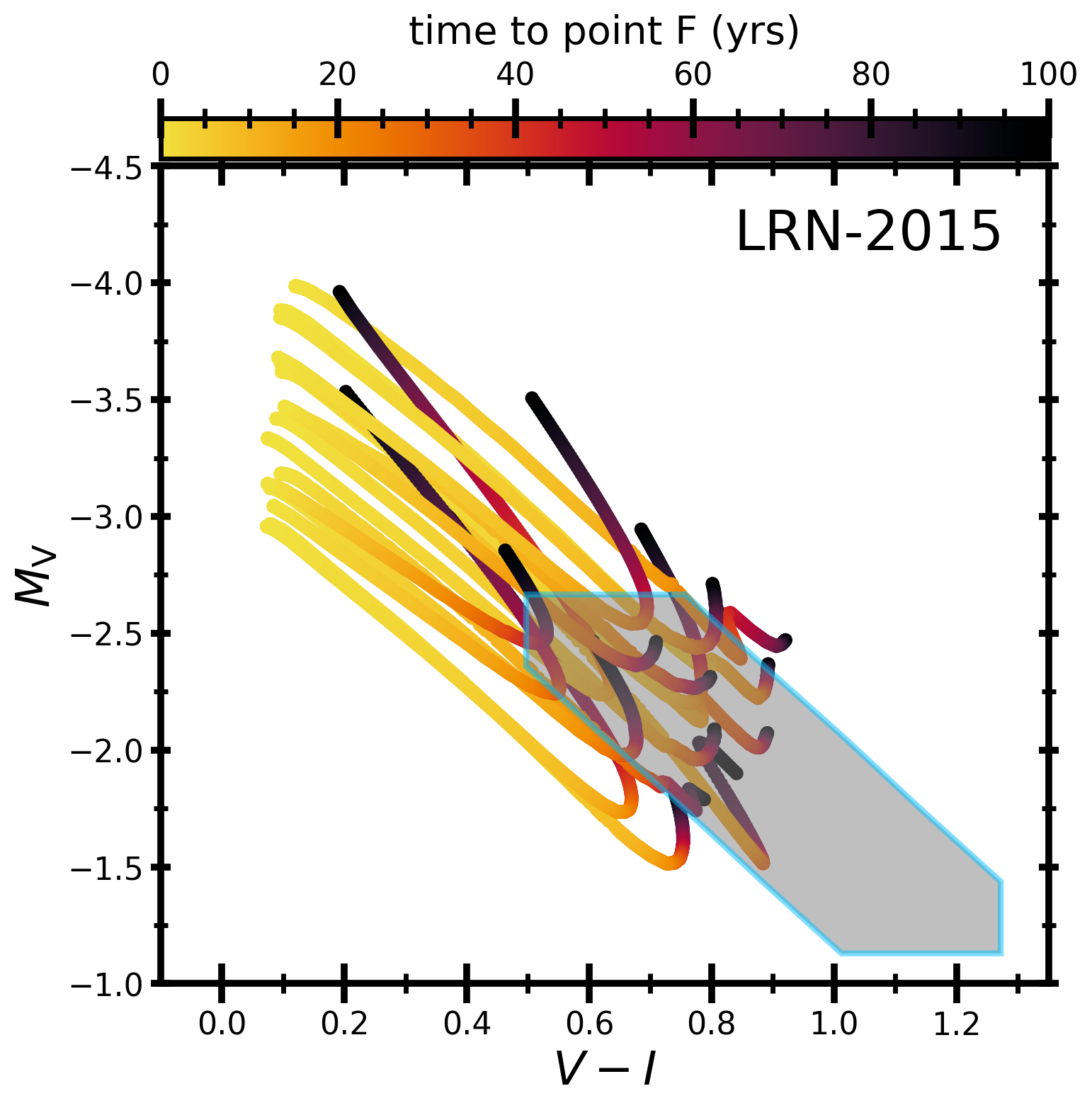}
	\caption{Colour-magnitude diagram track of DDI progenitor models for the 2015 LRN outburst in M31. The grey region with blue outline represents the observational uncertainties of the progenitor star, which include the distance modulus error, measurement error and extinction uncertainty. The coloured lines show the CMD tracks of matching DDI models (which include the approximated companion photometry) during the last century before point F. The colours represent the time from any given point on the track to point F as indicated in the colour bar.}
	\label{fig:M31_CMD}
\end{figure}

The LRN outburst in M31 was discovered in January 2015 by the MASTER network \citep{Shumkov2015}, about eight days before its peak brightness. At the location of the transient, a candidate progenitor star was identified in archival Hubble Space Telescope observations obtained in 2004 \citep{Dong2015, Williams2015}. Data obtained from the iPTF time-domain survey \citep{Rau2009a} showed a steady increase in the progenitor's luminosity for two years before the outburst \cite{Blagorodnova2020}. These archival observations provided insight into the pre-outburst photometric evolution, making it possible to infer key properties of the presumed donor star, such as mass, radius and age at the moment of RLOF. 

We compare our models to the photometric properties reported in \cite{MacLeod2017} from roughly 10.5 yrs before the outburst, which were derived from \cite{Williams2015} by varying the amount of extinction within the reported observational uncertainties. Using single-star models to match the observed progenitor photometry, \cite{MacLeod2017} found a good match for a $3-5.5\;M_{\odot}$ donor star filling its Roche lobe at a radius of roughly $30\; R_{\odot}$. As discussed in Section \ref{sec:discussion}, our models cannot accurately predict the time of the outburst and it may occur before point F is reached. We thus consider models of donor stars that are within one century of reaching their point F. We use the bolometric corrections from \cite{Choi2016} to calculate V and I magnitudes from our binary models. Similarly to the estimate in Section \ref{sec:discussion_accretor}, we compute an estimate of the V and I magnitudes of the accretors by interpolating in MIST models for ZAMS stars. We then calculate the total magnitudes of our binaries as
\begin{align}
    M_{\rm V, tot} = -2.5 \log \left( 10^{-M_{\rm V, d}/2.5} + 10^{-M_{\rm V, a}/2.5}  \right),
\end{align}
and analogously for the I band magnitude. The results of this comparison are presented in Figure \ref{fig:M31_CMD}.

We infer from our models that match the observed progenitor within the observational uncertainties during their rapid brightening phase a possible progenitor donor mass of $6-10\;M_{\odot}$, with a donor star that fills its Roche lobe on the mid-to-late HG (at a radius of about $30-60\; R_{\odot}$). As such, we find progenitor stars that are substantially more massive and more evolved than based on single-star models. A similar conclusion was reached for AT 2018bwo in NGC 45 by \cite{Blagorodnova2021}. 

Based on our calculations, the donor stars that fall within the observational uncertainties would evolve to point F in at least about 4~yrs and at most the assumed maximum of 100~yrs, depending on the initial donor mass and binary configuration. We note that all binary models in Figure \ref{fig:M31_CMD} match the observations before their luminosity minimum (point D). If, instead, we limit our selection to those models that increase in luminosity, we find that the amount of time to point F decreases to between 15 and 70 yrs. Assuming the outburst occurs at or before point F, this is roughly consistent with the observations.

\subsection{Observational signatures of (un-)stable mass transfer}
\label{sec:observational_signatures}

Donor stars that undergo RLOF while crossing the HG may exhibit substantial and systematic changes in their luminosities (Sections \ref{sec:representative_evolution} and \ref{sec:parameter_space}). Photometric surveys with long enough baselines (see Section \ref{sec:gaia_observability}) may be able to observe stars exhibiting such changes. Here, we explore to what extent the photometric changes of the donor stars predicted by our models are potentially detectable in the forthcoming data releases of the Gaia mission for stars in the Milky Way. Additionally, we want to learn if such observations could be used to distinguish between DDI and stable mass-transfer binaries.

\subsubsection{Synthetic population of mass-transferring transients}
 We construct a synthetic population from our binary evolution models by assuming that the properties of the primordial binary population are distributed as follows. The initial donor masses are drawn from a Kroupa distribution \citep{Kroupa1993} between 0.1 $M_{\odot}$ and 100 $M_{\odot}$, the initial mass-ratios are distributed uniformly between $M_{\rm a}/M_{\rm d} = 0$ and $M_{\rm a}/M_{\rm d} = 1$, and the initial semi-major axes are distributed uniformly in $\log a/R_{\odot}$ between $a = 1 R_{\odot}$ and $a = 10^6 R_{\odot}$ \citep[but see e.g.][]{Moe2017}. 

As discussed in Section \ref{sec:discussion_accretor}, the evolution of the accretors could greatly affect observational properties of the binaries we simulated. As the accretor stars gain mass, they will increase in radius and luminosity, potentially outshining their donor star companions. Analogous to the method in Section \ref{sec:discussion_accretor}, we compute an estimate of the accretor luminosity $L_{\rm a}$ by interpolating in MIST models.

The inclusion of the accretor luminosity affects the stable mass-transfer binaries the most because those binaries start out with more luminous companions. Additionally, the overall change in the accretor mass and hence luminosity is greater there. For such binaries, we find that the accretor luminosity would typically overtake that of their companion donor stars well before the donor luminosity minimum (point D) has been reached. At point D, the donor stars account for $10-97 \%$ of total luminosity of the binaries ($L_{\rm tot}$), with a weighted average of $62\%$. In the following, we will use the label `minL' to refer to the point where $L_{\rm tot}$ reaches its minimum. For the DDI binaries, we find that the donor stars are always responsible for at least 73\% of the total luminosity of the binaries (with a weighted average of 97\%) and can thus be expected to remain the dominant feature in observed photometry during the entire slow dimming phase and the rapid brightening phase that follows it, barring the caveats outlined below. For these binaries, point D thus coincides with the minL point.

There are important caveats for the estimates of photometric change rates that follow. Firstly, our treatment of the accreting companion stars as ZAMS stars may significantly underestimate their luminosities. It is generally expected for an accreting MS star to increase substantially in luminosity if it receives mass on a timescale shorter than its thermal timescale \citep[see e.g.][]{Kippenhahn1977,Zhao2024}. On the other hand, the possibility of (highly) non-conservative mass transfer might result in a lower accretor luminosity. Secondly, we do not consider models where $\left . M_{\rm a}/M_{\rm d}\right|_{\rm RLOF} < 0.1$. This is mainly due to the expected strong effect of tides, which would make our predictions for such binary systems unreliable \citepalias[see also the discussion in][]{Temmink2023}. Lastly, we only take into account the first phase of mass transfer. In reality, many of the first-phase accretor stars will fill their Roche lobes in a second phase of mass transfer and potentially create another wave of transients. Hence, the total number of DDI systems from this mass range is most likely larger than estimated here, but it is uncertain how the ratios of DDI-to-stable mass transfer systems would be affected. In our setup, we find about 5.6 binaries undergoing stable mass transfer for every DDI binary.

\subsubsection{The phase of slow dimming}
\label{sec:slow_dimming_stats}

\begin{figure}[h!]
	\centering
    \includegraphics[width=\hsize]{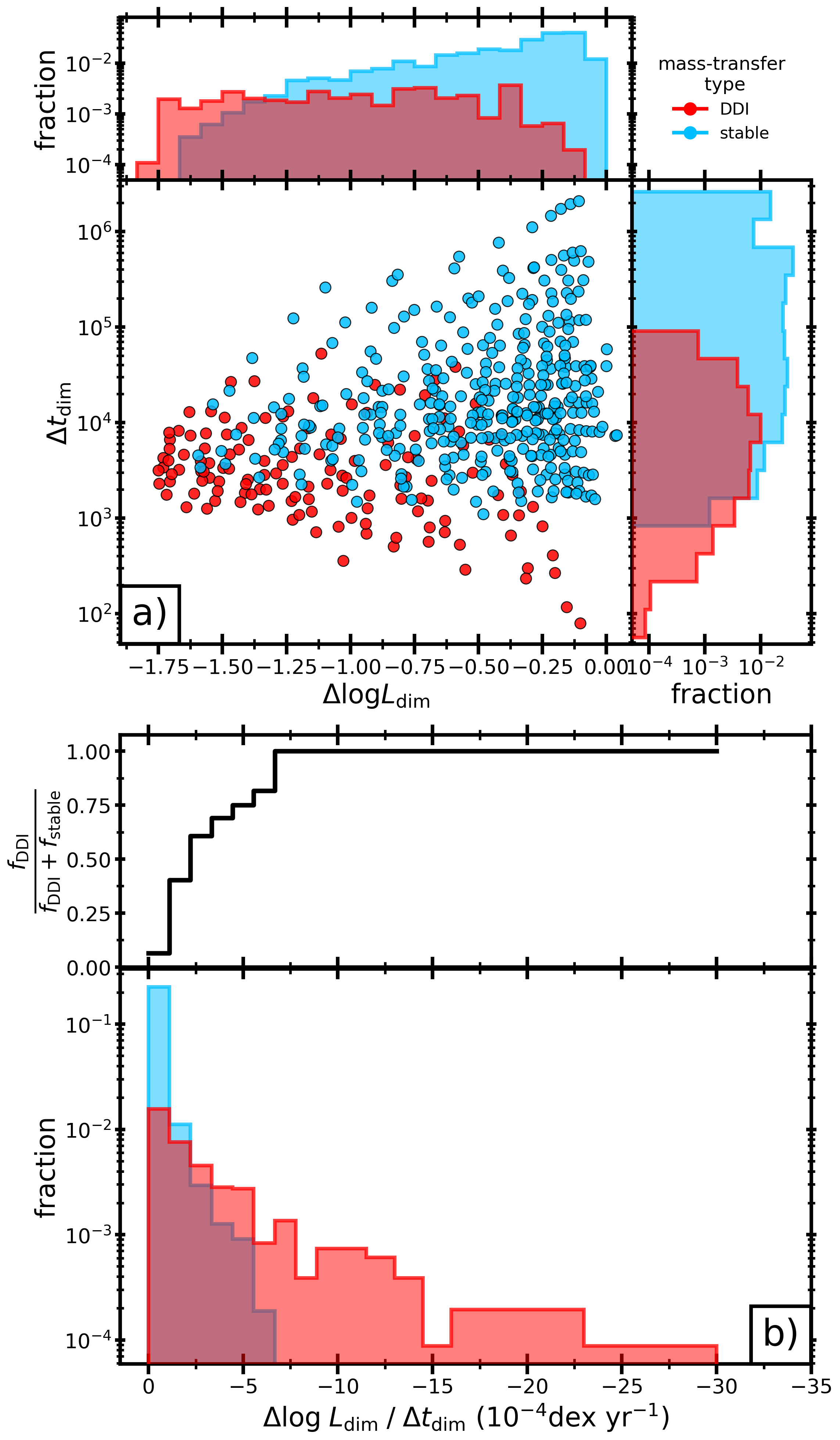}
	\caption{Photometric evolution of the binaries (donor and accretor luminosities combined) in our synthetic sample during their phase of slow dimming. Properties relating to DDI binaries are shown in red and those relating to stable mass-transfer binaries are shown in blue, as in the legend. The `fraction' label on the distributions refers to a fraction out of all stars with $2.5 \leq M_{\rm d}/M_{\odot} \leq 10$. Subfigure a shows the change in luminosity during this phase plotted against the corresponding dimming evolution time. Subfigure b shows the distribution of the resulting dimming rates in the bottom panel, and the ratio of DDI binaries to all binaries as a function of dimming rate in the top panel.}
	\label{fig:dimming_brightening_stats}
\end{figure}

The phase of slow dimming is a characteristic of mass transfer that is shared by both the DDI and stable mass-transfer binaries (see Section \ref{sec:representative_evolution}). During this phase, the donor stars become up to a hundred times less luminous than they were at the onset of RLOF. For the DDI binaries, the evolution time between points B and minL ($\Delta t_{\rm dim}$) is up to about fifty thousand yrs, whilst for the stable mass-transfer binaries --which have less extreme initial mass ratios-- this phase can last for up to about two million years.

Despite these extremely long timescales and the apparent similarity of this phase between DDI and stable mass-transfer binaries in terms of their HRD evolution, we show that observations of binaries in this phase may potentially be used to infer the final fate (stable vs DDI) of the binaries. In the following, we estimate the (relative) probability that a given observed dimming binary will ultimately experience a mass-transfer instability, based on the measured rate of dimming.

Figure \ref{fig:dimming_brightening_stats}a shows that both binary types span similar ranges in luminosity difference, $\Delta \log L_{\rm dim} = \log L_{\rm tot, minL} - \log L_{\rm tot, B} \approx -1.75$ to $0$. However, their distributions differ: stable mass transfer binaries are skewed toward smaller luminosity drops. Their weighted mean $\Delta \log L_{\rm dim}$ is about $-0.5$, compared to $-1$ for DDI binaries. The biggest contrast lies in the $\Delta t_{\rm dim}$ distributions. Stable mass transfer binaries show a broad range—from 1 to 2,100 kyrs (mean: 270 kyrs)—while DDI binaries have much shorter timescales, from 80 to 53,000 yrs (mean: 10,000 yrs).

Figure \ref{fig:dimming_brightening_stats}b shows the rates of dimming of our binaries, calculated simply as $\Delta \log L_{\rm dim}/\Delta t_{\rm dim}$. The relatively longer evolution time and smaller change in total binary luminosity result in a concentration of stable mass transfer models towards the lowest rates of dimming. The DDI binaries on the other hand, span a large range of dimming rates. Consequently, the relative proportion of DDI binaries increases sharply with the (absolute) dimming rate, as can be seen in the top panel of Figure \ref{fig:dimming_brightening_stats}b. Most notably, our results suggest that every binary with a primary star in the initial mass range $2.5 - 10 \;M_{\odot}$ dimming at a higher rate than $-2\cdot 10^{-4}$~dex yr$^{-1}$ (equivalent to about $\sim 5\cdot 10^{-4}$~mag yr$^{-1}$) is likely a DDI binary, and any such binary dimming faster than $-7\cdot 10^{-4}$~dex yr$^{-1}$ is a pre-DDI system. As such, our results could provide a powerful tool for detecting pre-outburst binaries, and be used to create efficient selection criteria for LRNe progenitor searches, provided the observational baseline is long enough (see Section \ref{sec:gaia_observability}).

\subsubsection{The phase of rapid brightening}
\label{sec:rapid_brightening_stats}

\begin{figure}[h!]
	\centering
    \includegraphics[width=\hsize]{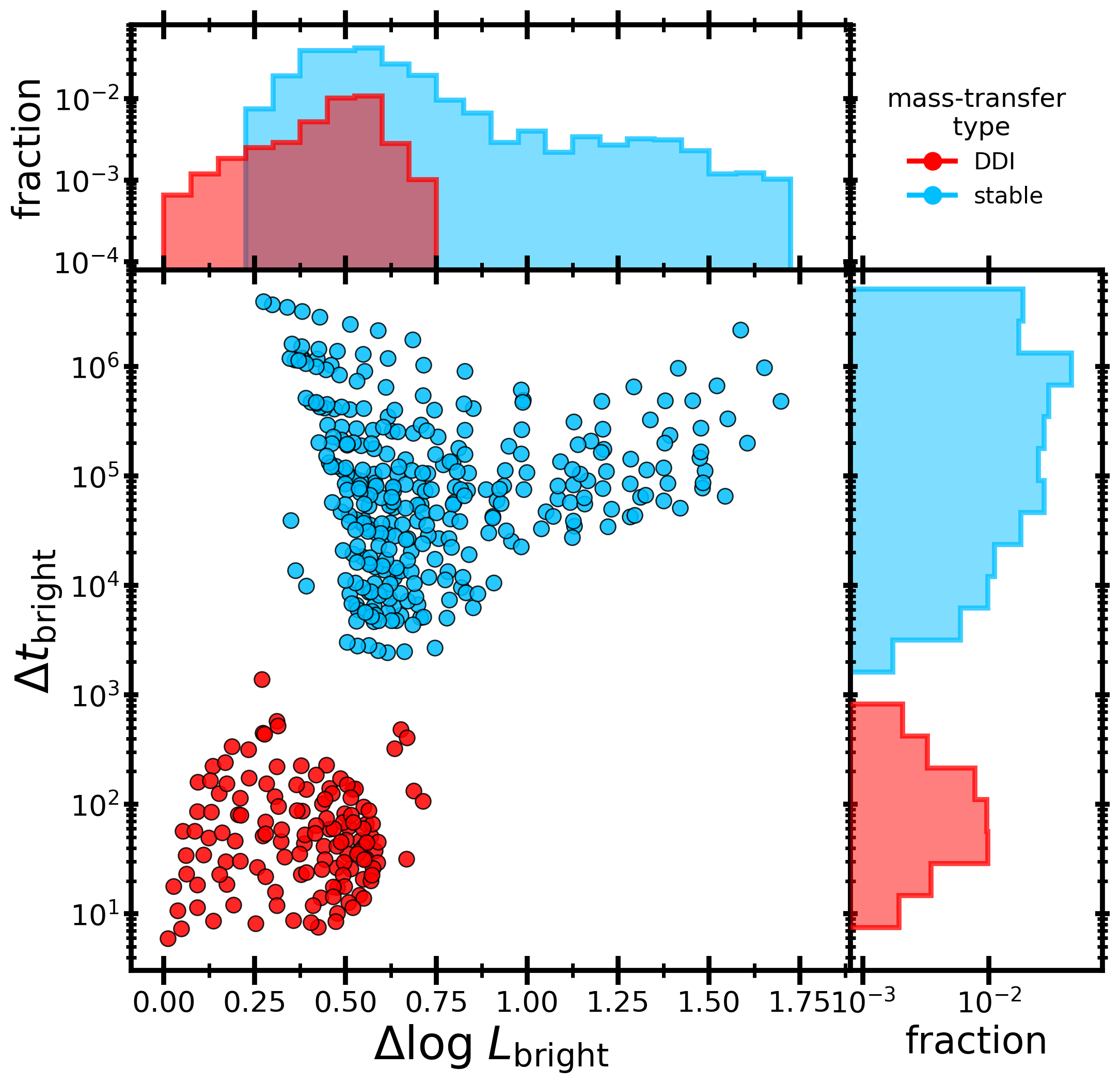}
	\caption{Photometric evolution of the binaries (donor and accretor luminosities combined) in our synthetic sample during their phase of rapid brightening, analogous to Figure \ref{fig:dimming_brightening_stats}a.}
	\label{fig:brightening_stats}
\end{figure}

The rapid brightening phase directly following the slow dimming phase could also provide valuable insights for the DDI binaries. With a weighted average brightening time ($\Delta t_{\rm bright}$) of about 130~yrs, this part of the evolution typically occurs on a `human' timescale, making it particularly suitable for observations. The full range of evolution times is about $6~{\rm yrs}\leq\Delta t_{\rm bright}\leq 1,400~{\rm yrs}$, as can be seen in Figure \ref{fig:brightening_stats}. During this phase, the luminosity increases by a factor between one and five, with a weighted average of nearly three.

Figure \ref{fig:brightening_stats} also shows statistics for the stable mass-transfer binaries. Both $\Delta t_{\rm bright}$ and  $\Delta \log L_{\rm bright}$ span a significantly larger range in values for these binaries than for the DDI binaries. More explicitly, $3,6 {\rm kyrs}\leq\Delta t_{\rm bright}\leq 3,9 {\rm Myrs}$ with a weighted mean of 790 kyrs and $0.25 \leq\log L_{\rm bright}\leq 1.75$  with a weighted mean of 0.4. Whilst the typical relative increase in luminosity is comparable between the two sets of binaries, the timescales over which this happens are clearly vastly different.

For the stable mass-transfer binaries, the average rate of brightening ($\Delta \log L_{\rm bright}/\Delta t_{\rm bright}$) is approximately $2\cdot 10^{-4}$~dex yr$^{-1}$, which is about ten times faster than their average dimming rate and comparable to those of single stars of similar luminosities ascending the giant branch. As such, brightening at this rate --if observable-- cannot be used as a sign of active mass transfer. By contrast, the DDI donor stars increase in luminosity at an average rate of about $9\cdot 10^{-3}$~dex yr$^{-1}$, which is about 45 times more rapid than the stable mass-transfer binaries.

\subsubsection{Implications for Gaia}
\label{sec:gaia_observability}

\begin{figure*}[h!]
	\centering
	\includegraphics[width=\hsize]{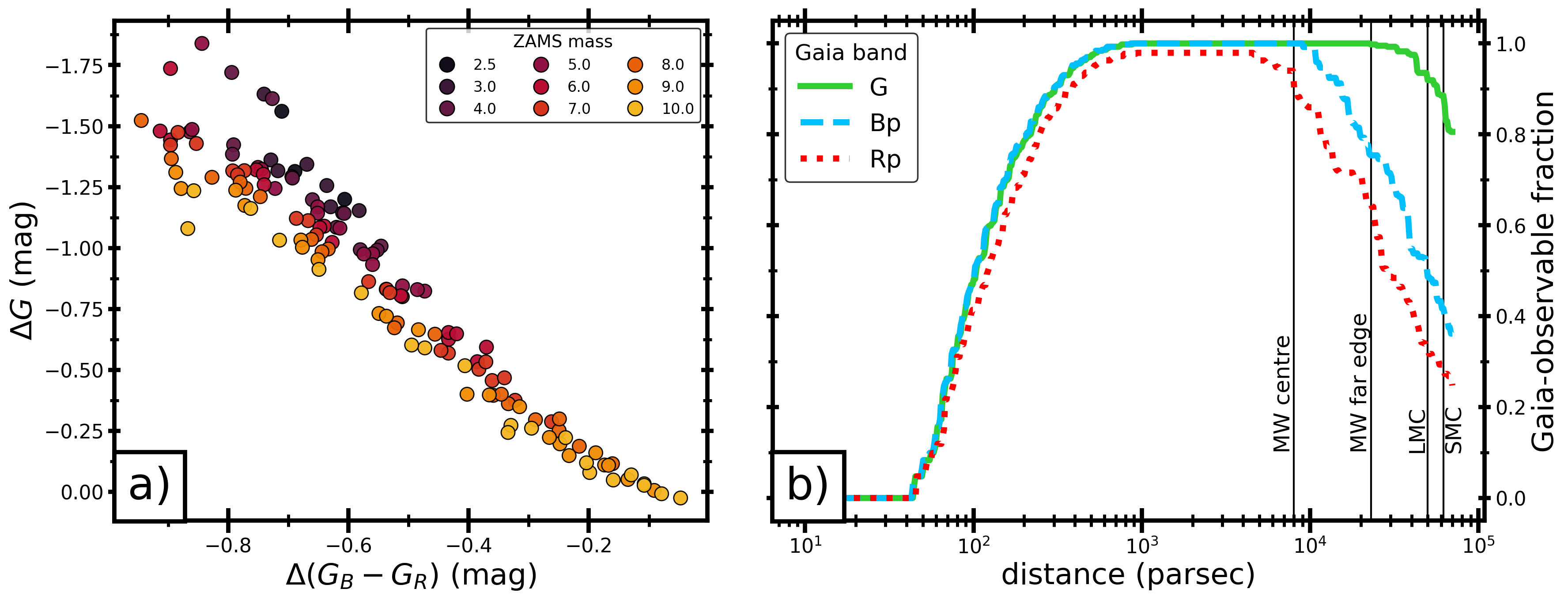}
	\caption{Chromatic evolution of the DDI binaries during their phase of rapid brightening in the Gaia bands. Panel a shows the the change in Gaia magnitude and colour during this phase, with the origin representing the point of minimum luminosity. The colours denote the ZAMS mass of the donors, as in the legend. Panel b shows the fraction of DDI binaries for which photometric changes are expected to be detectable by Gaia as a function of distance --neglecting effects of extinction-- in each Gaia photometric band, coloured accordingly as in the legend. The vertical black lines indicate, from left to right, the approximate distance to the galactic centre, the far edge of the Milky Way (based on its $D_{25}$-isophotal diameter \citet{Goodwin1998}), and the large and small Magellanic clouds.}
	\label{fig:brightening_stats_gaia}
\end{figure*}

With more than three trillion observations of two billion stars spanning over one decade, Gaia is very well-suited for potentially detecting the photometric changes associated with mass transfer. We focus on the brightening phase in the DDI binaries, because the photometric evolution is the most significant there. As before, we use bolometric corrections from \cite{Choi2016} to obtain Gaia magnitudes for our model binaries. The change in Gaia CMD position of the DDI donor stars during the rapid brightening phase is shown in Figure \ref{fig:brightening_stats_gaia}a.

It is clear that the evolution during this phase can be quite substantial. The DDI donors exhibit up to about 1.75 magnitudes of brightening (with a weighted average of one magnitude) and up to a full magnitude of shifting towards the blue ($G_{\rm B} - G_{\rm R}$ decreases by 0.6 magnitudes, on average). The brightness and colour of the point of minimum total luminosity, the starting point of the rapid brightening phase, depend on the initial mass ratio (see Section \ref{sec:parameter_space}). However, the relative evolution from that point onwards appears fairly independent thereof. We furthermore find that the extent of the luminosity change for donors of initially higher masses spans a wider range, and extends to almost vanishing brightness increase, whilst donors of initially lower mass tend to concentrate more at the upper end of the range of brightness increase.

To estimate the detectability of photometric changes due to mass transfer, we compare the photometric evolution of DDI binaries to the Gaia photometric uncertainties. We estimate the latter by using the tool provided by Gaia DPAC\footnote{Available here: \href{https://www.cosmos.esa.int/web/gaia/fitted-dr3-photometric-uncertainties-tool }{https://www.cosmos.esa.int/web/gaia/fitted-dr3-photometric-uncertainties-tool}.} to reproduce the Gaia (E)DR3 photometric uncertainties described in the GAIA-C5-TN-UB-JMC-031 technical note using data in \cite{Riello2021}. We vary the distance between 10 and 70,000 parsec and convert our absolute Gaia magnitudes into apparent magnitudes for the three Gaia bands. We assume a DDI binary can be detected at a given distance if its average change in apparent magnitude over one year is larger than the corresponding Gaia uncertainty at that magnitude. For the purpose of this exercise, we ignore extinction and reddening, though we note that these could heavily influence our results.

The results of this exercise are shown in Figure \ref{fig:brightening_stats_gaia}b. Up to about 1 kpc the fraction of DDI binaries whose photometric changes are detectable by Gaia increases to nearly 100\%. After a sizeable plateau, this fraction begins to drop off for distances larger than about 10kpc to 30kpc, depending on the specific Gaia band. The Gaia-detectable fraction remains above 50\% out to 26 kpc, 49 kpc and 70 kpc, for the Gaia R, B and G bands respectively. We thus find that, for binaries with a primary star in the initial mass range $2.5 - 10 \;M_{\odot}$ that are observable by Gaia, photometric changes associated with pre-DDI evolution should be detectable throughout the entire Milky Way and potentially out to the Magellanic Clouds.

\section{Conclusions}
\label{sec:conclusion}

In this work, we have systematically studied the pre-instability evolution of binary systems with a HG donor of initial mass $2.5 - 10\; M_{\odot}$. We found that all donor stars studied here go through a qualitatively similar evolution, consisting of a slow dimming phase followed by a phase of rapid brightening. The phase of brightening is very short-lived (lasting only a few years to a few centuries) and coupled with a strong increase in effective temperature. We find that the strong increase in surface luminosity during this phase is due to the energy released by the ever more rapid recombination of ionized hydrogen in the outermost surface layers due to the rapid mass loss. Donor stars experiencing stable mass transfer are known to also experience phases of dimming and brightening. However, their brightening phase is orders of magnitude slower (occurring due to the natural expansion of the donor star on its nuclear timescale) and paired with a decrease in effective temperature. 

Our models are subject to several theoretical limitations, most notably that we did not evolve the accreting companion star and that we treated the mass transfer as conservative. To investigate the potential effect of non-conservative mass-transfer, we re-simulated several of our DDI binaries with varying mass transfer efficiencies, finding that the qualitative picture of pre-DDI evolution still holds. However, non-conservative mass transfer could lead to outflows and dust formation that would impact the observable aspects of our predictions. Our models could be further improved in future work by accounting for stellar rotation and tidal interactions.

We have applied our models to observations of the progenitor stars of two LRN outbursts. For the outburst in V838 Mon, we identified a previously unexplored progenitor scenario that involves unstable mass transfer from a post-MS star. For the 2015 outburst in M31, we found that a DDI progenitor donor star could have a significantly higher mass than in previous studies that did not take into account detailed pre-instability evolution \citep[e.g.][]{MacLeod2017}, whilst remaining consistent with the pre-outburst photometry.

We found that it may be possible to identify whether mass transfer in a particular binary system will remain stable or become unstable based on its rate of luminosity change. This might prove a powerful tool for the exploration of large observational data sets and for helping to efficiently target a specific parameter space for future observations. Additionally, based on our models, we expect the photometric change of a significant fraction of the pre-DDI binaries in their rapid brightening phase to be detectable by Gaia throughout the Milky Way. Several searches for slowly varying sources have already been carried out \citep[see e.g.][]{Addison2022, Petz2025}. Although extinction currently poses a major obstacle for high-confidence candidate selection, this will improve with future data and dust map releases.

This study of the delayed dynamical instability in mass-transferring binaries fills an important gap in our understanding of how luminous red novae and similar transients arise. While unstable mass transfer is a well-accepted pathway to these events, the long-lived pre-instability phase in systems with Hertzsprung gap donors has thus far received little attention. By capturing the full progression leading up to the DDI, this work goes beyond previous case studies that focused mostly on the onset or aftermath of instability. By incorporating this phase, this work offers a more complete framework for connecting theoretical models to transient observations and opens the door to more accurate classification and prediction of mass-transfer-driven events.

\begin{acknowledgements}
    We thank the anonymous referee for their constructive remarks.
	KDT acknowledges support from NOVA.
    SJ thanks Ulrich Kolb and Philipp Podsiadlowski for discussions regarding the delayed dynamical instability.
\end{acknowledgements}

\bibliographystyle{aa}
\bibliography{DDI-LRNe}

\begin{appendix}

\section{Common envelope evolution}
\label{sec:CEE}

\begin{figure}[t!]
	\centering
	\includegraphics[width=\hsize]{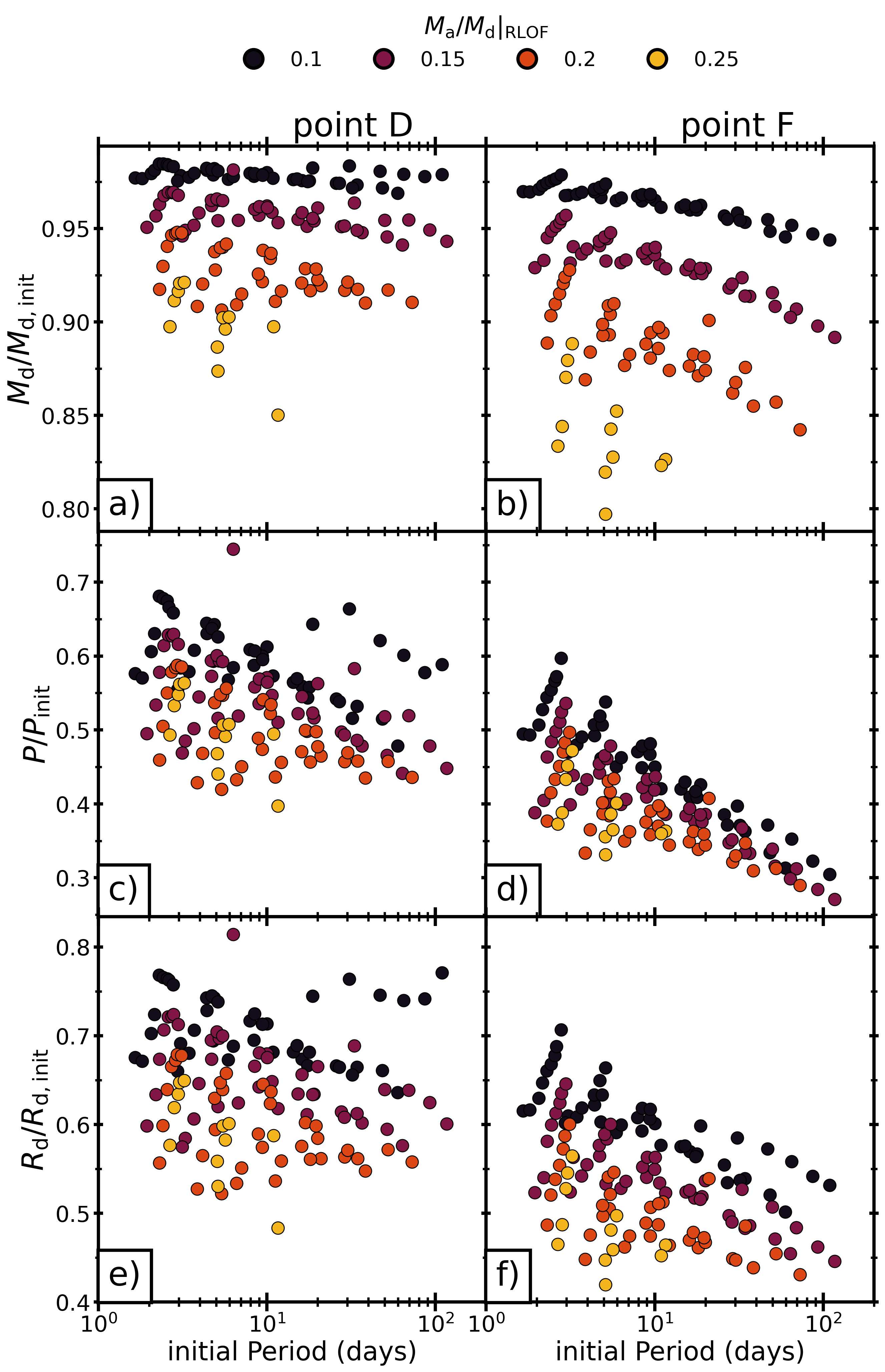}
	\caption{ Properties of DDI donor stars at points D (left column) and F (right column), plotted against their initial period. Panels a and b show the remaining mass relative to the total initial donor mass, panels c and d show the ratio of the period to the initial period and panels e and f show the donor star radius relative to its radius at RLOF. In all panels, the data points are coloured according to the initial mass ratio of the binary, as indicated in the legend.}
	\label{fig:variation_DF}
\end{figure}

\begin{figure}[t!]
	\centering
	\includegraphics[width=\hsize]{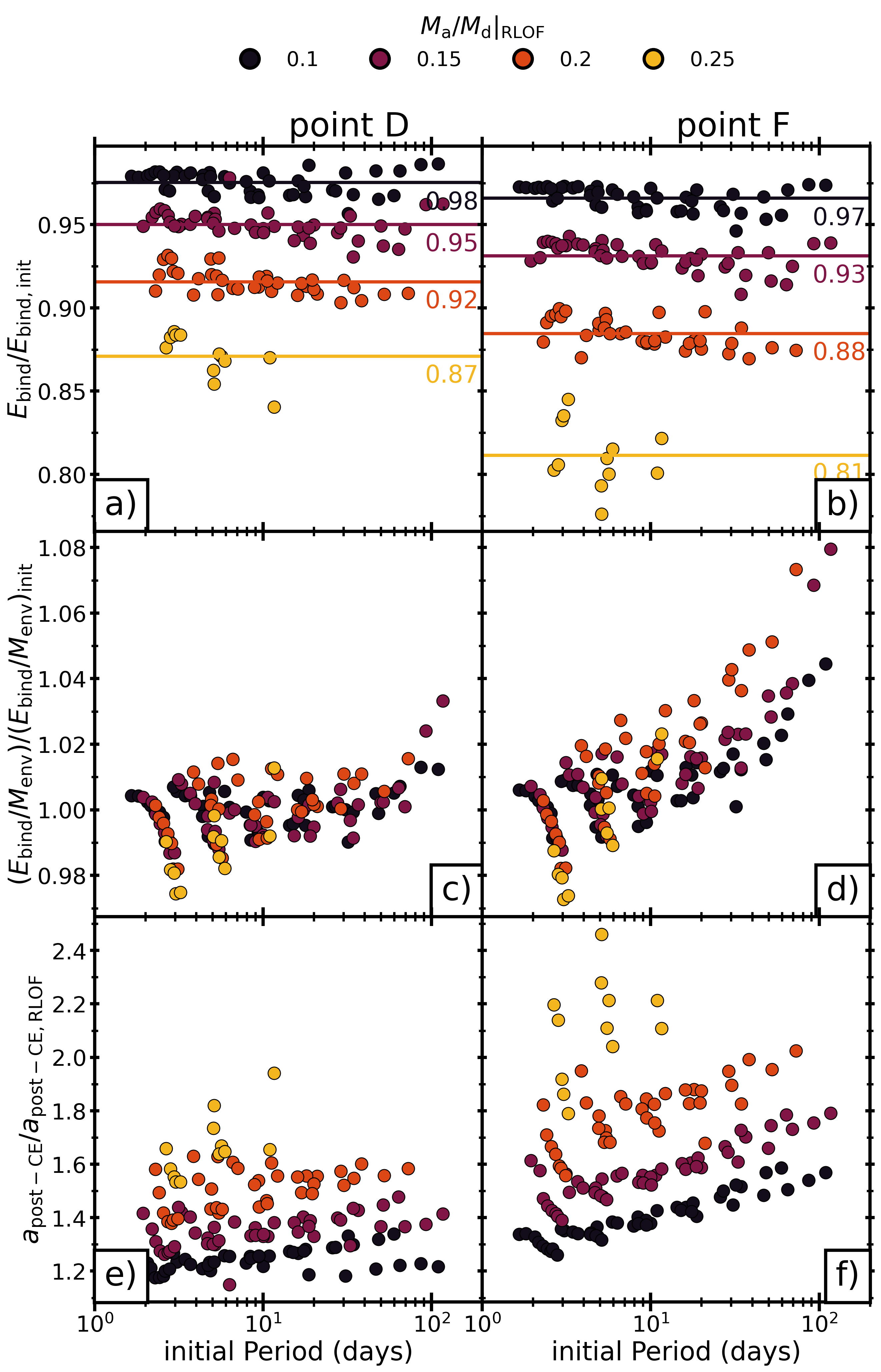}
	\caption{Similar to Figure \ref{fig:variation_DF}. Panels a and b show the binding energy of the envelope of the donor stars relative to the onset of RLOF, panels d and d show the specific binding energy relative to the onset of RLOF and panels e and f show the post-CE orbital separation calculated using binary properties at points D and F relative to to the post-CE separation calculated using the properties at RLOF.}
	\label{fig:CE_stats}
\end{figure}

One of the consequences of dynamically unstable mass transfer is a CE phase, during which the envelope of the donor star engulfs both binary component stars. Amongst the many challenges involved with the modelling of a CE is taking into account the pre-instability evolution in an appropriate manner. As explored in the main text of this work, this pre-instability evolution can be rather extended, leading to substantial differences in the observational appearance of the pre-CE donor star, as well as the initial conditions of a potential CE phase compared to the onset of RLOF. 

As discussed in Section \ref{sec:discussion_accretor}, a CE can potentially occur between points D and F. Figure \ref{fig:variation_DF} shows the change in the masses and radii of our donor stars, as well as the change in the orbital periods, by the time points D and F are reached. The donor stars can lose between 2 and 20 per cent of their initial total mass, as shown in Figures \ref{fig:variation_DF}a and b. A larger fraction of mass is lost by donors in systems with initially less extreme initial mass ratios and initially larger periods.

Figures \ref{fig:variation_DF} c and d show the orbital periods at D and F relative to the initial orbital period. Donor stars in systems with initially less extreme mass ratios and larger initial periods are able to transfer more mass, and thus the orbits of such systems shrink more during the pre-DDI evolution. 

Figures \ref{fig:variation_DF} e and f show the donor radii at points D and F, relative to the donor radius at the onset of RLOF.  Up to point F, the radii of the donor stars only overfill their Roche lobes by up to about 5 per cent. Hence, the donor stars in systems with initially larger periods and less extreme mass ratios end up with smaller radii.

As a result of the changing structure of the donor stars, the binding energy of their envelopes decreases. We calculate the envelope binding energy as follows:
\begin{align}
	E_{\rm bind} = \int_{M_{\rm He}}^{M_{\rm d}} \left( e_{\rm int}(m) - G \frac{m}{r(m)}\right){\rm d}m
\end{align}
Here, $M_{\rm He}$ denotes the mass coordinate of the helium rich core boundary (defined as the outermost location in mass where $X<0.1$ and $Y>0.1$), $r(m)$ and $e_{\rm int}(m)$ are the radius and the total specific internal energy (including the potential energy release from ionization/dissociation) at mass coordinate $m$, respectively. The resulting binding energies at points D and F, relative to those at the onset of RLOF, are shown in panels a and b of Figure \ref{fig:CE_stats}. The fraction of remaining binding energy at these points is strongly dependent on the initial mass ratio, but almost independent of the initial donor mass or binary orbit. At both points D and F, the remaining binding energy closely tracks the remaining envelope mass (see Figure \ref{fig:CE_stats}c and d). This suggests that binding energy loss is primarily driven by mass loss, not by major structural changes in the donor stars.

In order to estimate whether or not these changes to the stellar structure could affect the outcome of a CE phase, we employ the so called $\alpha$-prescription \citep{Paczynski1976,Webbink1984, deKool1987,Livio1988,deKool1990} in which the energy budget of the binary system is used to estimate the outcome of a CE:
\begin{align}
	E_{\rm bind} = \alpha \Delta E_{\rm orb} = \alpha \frac{G M_{\rm a}}{2}\left( \frac{M_{\rm d,i} - M_{\rm env}}{a_{\rm post-CE}} - \frac{M_{\rm d, i}}{a_{\rm i}} \right)
\end{align}
where the subscripts i and post-CE indicate quantities evaluated at the initial and final moments of the CEE, $\Delta E_{\rm orb}$ is the change in binary orbital energy and $M_{\rm env}$ the mass of the donor envelope. We set $\alpha = 1$ and solve the equation for the post-CE semi-major axis $a_{\rm post-CE}$ (shown in Figure \ref{fig:CE_stats} e and f, relative to $a_{\rm post-CE}$ calculated from properties at the onset of RLOF). The change in binding energy leads to somewhat larger final separations, though always less than $1\; R_{\odot}$. We furthermore find that the companions nevertheless fill their Roche lobes, if we approximate the radii of the accretors by those of ZAMS stars using MIST models, as in Section \ref{sec:discussion_accretor}. It is thus unlikely that a CE phase could be survived by our binaries, unless the effective value of $\alpha$ is significantly greater than we assume. Nevertheless, our results might be of use in detailed CE modelling studies, where our models could be used as initial conditions.

\end{appendix}

\end{document}